\documentstyle[12pt]{article}

\begin{document}
\begin{titlepage}

\vspace {1in}

\begin{center}
{\Large\bf Simulations with Complex Measure}\\

\vspace{0.5in}

J F Markham$^a$ and T D  Kieu$^{a,b}$\\
$^a$School of Physics, University of Melbourne, Parkville 3052,
Australia\\
$^b$Department of Physics, Columbia University, New York, NY 10027,
USA\\

\vspace{0.5in}

{\bf Abstract}
\end{center}
\begin{quotation}
A method is proposed to handle the sign problem 
in the simulation 
of systems having indefinite or complex-valued measures. In general, 
this new approach, which is based on renormalisation blocking, is shown 
to yield statistical errors smaller 
than the crude Monte Carlo method using absolute values of the 
original measures. The improved method is applied to the 2D Ising model 
with temperature generalised to take on complex values. It 
is also adapted to implement Monte Carlo Renormalisation Group 
calculations of the magnetic and thermal critical exponents.
\end{quotation}
\end{titlepage}
\section{The Monte Carlo sign problem}
In order to evaluate a multi-dimensional integral  
\begin{eqnarray}
I&\equiv& \int fdV
\label{1a}
\end{eqnarray}
using Monte Carlo (MC) one can sample the points in the integration domain 
with a non-uniform distribution, $p$, which reflects the contribution from 
the measure $f$ at each point, as in the {\it importance 
sampling}~\cite{BinderHeermann}. 
This sampling gives the following estimate for the integral:
\begin{eqnarray}
I &\approx& \left\langle 
\frac{f}{p}
\right\rangle \pm \sqrt{\frac{S}{N} }, 
\label{1}
\end{eqnarray}
where $N$ is the number of points sampled, $p\geq 0$ and is normalised
\begin{eqnarray}
\int p dV &=& 1,
\nonumber\\
\left\langle f/p \right\rangle&\equiv&\frac{1}{N}
\sum\limits_{i=1}^{N}f(x_{i} )/p(x_{i}).\nonumber
\end{eqnarray}
and
\begin{eqnarray}
S&\equiv& \int\left|\frac{f}{p}-I\right|^2 p dV, \nonumber\\
&\approx& \left\langle \frac{f^{2} }{p^{2} }
\right\rangle -\left\langle \frac{f}{p} \right\rangle ^{2}. 
\label{2}
\end{eqnarray}

The best choice of $p$ is the one that minimises the standard 
deviation squared $S$.  This can be found
by variational method leading to the {\it crude} average-sign 
MC weight~\cite{std}
\begin{eqnarray}
p_{\rm crude}&=&\frac{\left| f\right| }{\int \left| f\right| dV },
\label{3}
\end{eqnarray}
giving the optimal 
\begin{eqnarray}
S_{\rm crude} &=&\left( \int \left| f\right| dV \right) ^{2} -\left| 
\left( \int fdV \right) \right| ^{2}.
\label{4}
\end{eqnarray}

Applying to physical systems, MC method can be used to evaluate
the expectation value of some measurable quantity $\Theta$
\begin{eqnarray}
\left\langle \Theta\right\rangle& =&\frac{\sum\limits_{\{s\}}
\Theta e^{-H} 
}{\sum\limits_{\{s\}}e^{-H}}, \nonumber\\
{Z} &\equiv& \sum_{\{s\}} e^{-H},
\label{12}
\end{eqnarray}
where $s$ denotes the dynamical variables and $H$ is the hamiltonian of 
an equilibrium statistical system or
the action of some Euclidean quantum field theory.
  
Since an estimate of the denominator of~(\ref{12}) is independent of a 
particular measurable and is needed in calculating all observables, it is 
to this that we apply the MC method.
The integral~(\ref{1a}) now assumes the form of the partition function $Z$,
upon which the crude weight~(\ref{3}) takes on the explicit form
\begin{eqnarray}
p_{\rm crude}\quad &\rightarrow& \quad |e^{-H}| /\sum\limits_{\{s\}}|e^{-H}|.
\label{13}
\end{eqnarray}

The Boltzmann weight $e^{-H}$ can in general be real but non-definite, or
even be complex-valued, in which case it is possible to generalise the
absolute values of real numbers in the above expressions to those of
complex numbers.  And the error bars now can be visualised as the
{\it error radius} in the complex plane of a circle centred at the 
complex-valued central MC estimate.  The variational derivation still
goes through as with real numbers.

We then have in the MC approximation
\begin{eqnarray}
\left\langle \Theta\right\rangle &\approx&\frac{\sum\limits_{\rm MC 
configurations}\Theta\left(
e^{-H} /\left| e^{-H} \right|  \right) 
}{\sum\limits_{\rm MC configurations}\left( e^{-H} /\left| e^{-H} \right| 
\right)}
 \equiv \frac{\left\langle \left\langle \rm \Theta\right\rangle \right\rangle
 _{\rm crude}}
{\left\langle \left\langle \rm sign \right\rangle \right\rangle_{\rm crude}}.
\label{14}
\end{eqnarray}
In the above $\left\langle \left\langle \rm sign \right\rangle \right\rangle$
also denotes the average, with respect to a given MC weight, of the phase
when $e^{-H}$ is complex.

The {\it sign problem}~\cite{sign} arises when
$\left|\left\langle \left\langle \rm sign \right\rangle 
\right\rangle\right|$ is vanishingly small: then unless a huge number of
configurations are MC sampled, the large statistical fluctuations 
of~(\ref{14}), because of the small denominator, render the measurement 
meaningless.

Unfortunately, many interesting and important physical problems suffer the
sign problem like the real-time path integrals of quantum mechanics and quantum
field theory, lattice QCD at finite temperature and non-zero chemical potential,
lattice chiral gauge theory, quantum statistical system with fermions $\ldots$
None of the existing proposals is quite satisfactory: complex Langevin 
simulations~\cite{complex}
cannot be shown to converge to the desired distribution and often fail
to do so; others~\cite{others} are either restricted to too small a lattice,
too complicated, or not general enough or rather speculative.

In the next section we present another improved method, which is then applied to
the Ising model in two dimensions and the results will be compared with
the crude MC of this section, as well as with series-expansion data.
\section{The improved method}
One way of smoothing out the sign problem is to do part of the integral 
analytically, and the remainder using MC~\cite{kieu}.  The analytical summation
is not just directly over a subset of the dynamical variables; in
general it can be a renormalisation group (RG) blocking where coarse-grained 
variables are introduced.
We will show below 
that this does yield certain improvement over the crude MC in general.

Let $P\{V',V\}$ be the normalised RG weight relating the original variables 
$V$ to the blocked variables $V'$~\cite{huang},
\begin{eqnarray}
P\{V',V\} &\geq& 0,\nonumber\\
\int P\{V',V\} dV' &=& 1.
\nonumber
\end{eqnarray} 
Inserting this unity resolution into the integral~(\ref{1a})
\begin{eqnarray}
I &=& \int dV\int dV' P\{V',V\}f,\nonumber\\
&\equiv& \int dV' g(V'),
\label{5}
\end{eqnarray}
and assuming that the blocking can be done exactly or approximated to a good 
degree such that we then obtain $g$ as a function of blocked variables in closed
form. 
An example of the RG blocking which we will employ in the next section
for the Ising model
is the sum over spins on odd sites of the lattice, leaving behind a measure
$g$ in terms of the other half of the spins on even sites.
Thus, an MC estimator is only needed for the remaining integration over $V'$ 
in~(\ref{5}).  As with the crude method of the last section, variational
minimisation for $S$ of~(\ref{2}), with $g$ in place of $f$, leads to the 
{\it improved} MC
\begin{eqnarray}
p_{\rm improved}&=&\frac{|g|}{\int |g| dV'}.
\label{improved}
\end{eqnarray}

This one-step exact RG blocking already improves over the crude
average-sign method of the last section.  Firstly, the improved
weight sampling yields in~(\ref{14}) a denominator of magnitude not
less than that sampled by the crude weight:
\begin{eqnarray}
\left|\left\langle \left\langle \rm sign \right\rangle
\right\rangle_{\rm improved}\right| &\equiv& \frac{Z}{\int |g| dV'},
\nonumber\\
&=& \frac{Z}{\int\left|\int P\{V',V\}fdV\right|dV'}, \nonumber\\
&\geq& \frac{Z}{\int|f|dV},\nonumber\\
&\equiv&
\left|\left\langle \left\langle \rm sign \right\rangle
\right\rangle_{\rm crude}\right|,
\label{i1}
\end{eqnarray}
where we have used the definitions of the sampling weights in the second
equality, $g$ in~(\ref{5}).  The
inequality is the triangle inequality from the properties of $P$. 
In other words,
from their definitions, $\left\langle \left\langle \rm sign \right\rangle
\right\rangle_{\rm improved}$ is proportional to 
$\left\langle \left\langle \rm sign \right\rangle
\right\rangle_{\rm crude}$, with the proportionality constant is some
function of temperature and external fields.  Both of them vanish when
the partition function does; away from this point, however, the improved
method is no worse than the crude one.

Secondly, it is also not difficult to see that the statistical fluctuations 
associated with improved MC is not more than that of the crude MC,
\begin{eqnarray}
S_{\rm improved} - S_{\rm crude} &=& \int \left| \frac{g^2}{p_{\rm improved}}
\right|dV' - \int \left| \frac{f^2}{p_{\rm crude}}\right|dV,\nonumber\\
&=& \left( \int |g| dV'\right)^2 - \left( \int |f| dV\right)^2, \nonumber\\
&=& \left( \int \left|\int P\{V',V\}f dV\right| dV'\right)^2 - \left( \int 
|f| dV\right)^2,
\nonumber\\
&\leq& 0,
\label{i2}
\end{eqnarray}
where we have used the definitions of the sampling weights in the second
equality, definition of $g$~(\ref{5}) in the third.  The
last inequality is the triangle inequality from the properties of $P$. 

Thus the RG blocking always reduces, sometimes significantly, the 
statistical fluctuations of
an observable measurement by reducing the magnitude of
\[ \frac{\sqrt{S}}{\left|\left\langle \left\langle \rm sign \right\rangle
\right\rangle\right|}.\]

Note that the special 
case of equality in~(\ref{i1},\ref{i2}) occurs {\it iff} there was no sign 
problem to begin with.
How much improvement one can get out of the new MC weight, i.e. how large
are the above inequalities, depends on the
details of the RG blocking and on the original measure $f$.
\section{Application to the 2D Ising model}
The Hamiltonian for the Ising model on a square lattice is
\begin{eqnarray}
H&=&-j\sum\limits_{\left\langle nn\prime \right\rangle }s_{n} s_{n\prime }
-h\sum\limits_{n}s_{n}.
\label{11}
\end{eqnarray}
Here we allow $j$ and $h$ to take on complex values in general.
The sum over $\{s\}$ in the partition function is a sum over all possible 
values of the spins $s_n  = \{+1,-1\}$ at site $n$. The sum over 
$\langle n'n\rangle$ is a sum over all nearest neighbours on the lattice. 
For the finite lattice, periodic boundary conditions are used.

The phase boundaries for the complex temperature 2D Ising model 
with $h=0$ are found by~\cite{shrock}
\begin{eqnarray}
{\rm Re} \left( u\right) &=&1+2^{\frac{3}{2} } \cos \omega +2\cos
2\omega  \nonumber \\
{\rm Im} \left( u\right) &=&2^{\frac{3}{2} } \sin \omega +2\sin
2\omega  
\end{eqnarray}
where $\omega $ is taken over the range $0\leq \omega \leq 2\pi$,
and
\begin{eqnarray}
u=e^{-4j}.
\end{eqnarray}
In the $ u$ plane, this is a limacon, which transforms to the 
$ j$ plane as shown in Fig. 1.  

\begin{center}
{{\bf Figure 1:} Phase diagram in the complex $j$
plane, $h=0$.}\\
{FM=ferromagnetic, PM=paramagnetic, AFM=antiferromagnetic.}
\end{center}
As $e^{-H} /\left| e^{-H} \right|$ takes values
on the unit circle, the crude MC estimator for the denominator of~(\ref{14})
might be vanishingly small, but its standard variance $S$ is of order unity,
leading to the sign problem.

For the improved method,
we adopt a simple RG blocking over the odd sites, labeled $\circ$.
That is,
the analytic summation is done over the configuration space spanned 
by the $\circ$ sites; while MC is used to evaluate the sum over the remaining 
lattice of the $\bullet$ sites. The following diagram shows the two 
sublattices, and 
how the $\bullet$ sites are to be labelled relative to the $\circ$ sites, 
for the site labelled $x$.
\begin{center}
{{\bf Figure 2:} Relative spin positions on a
partitioned lattice.}
\end{center}
In general, with finite-range interactions between the spins, one can
always subdivide the lattice into sublattices, on each of which the spins
are independent and thus the partial sum over these spins could be carried 
out exactly.

Summing over the spins $s_\circ$,
\begin{eqnarray}
Z
&=&\sum\limits_{\{s_\bullet\}} e^{h\sum\limits_{\bullet sites} s_{\bullet } } 
\prod\limits_{\circ sites}2\cosh \left[ js_{\bullet }^{+} +h\right]   
\end{eqnarray}
where $s_{\bullet }^{+} \equiv s_{\bullet }^{\uparrow } +s_{\bullet
}^{\rightarrow } +s_{\bullet }^{\downarrow } +s_{\bullet }^{\leftarrow } $.
The improved MC weight is then the absolute value of the summand on
the right hand side of the last expression for $Z$.

The quantities to be measured are magnetisation, $M$, and susceptibility,
$\chi$. 
These can be expressed in terms of the first and second derivatives of $Z$ 
respectively, 
evaluated at $h=0$.  Using the above notation:
\begin{eqnarray}
\frac{\partial Z}{\partial h} &=&\sum\limits_{\bullet spins}\left\{ 
\left[ e^{h\sum\limits_{\bullet sites}s_{\bullet }  } \prod\limits_{\circ sites}
2\cosh \left[ js_{\bullet }^{+} +h\right]  \right] \right.\nonumber\\
&&\left.\left[
\sum\limits_{\circ sites}\left( s_{\bullet }^{\uparrow } +\tanh 
\left( js_{\bullet }^{+}
+h\right) \right)  \right] \right\},
\end{eqnarray}
and
\begin{eqnarray}
\frac{\partial ^{2} Z}{\partial h^{2} } &=&\sum\limits_{\bullet spins}
\left\{ \left[ e^{h\sum\limits_{\bullet sites}
s_{\bullet }  } \prod\limits_{\circ sites}
2\cosh \left[ js_{\bullet }^{+} +h  \right] \right]\right. \times\nonumber\\
&&\left.
\left[ \left( \sum\limits_{\circ sites}
\left( s_{\bullet }^{\uparrow } +\tanh \left( js_{\bullet }^{+}
+h\right) \right)  \right) ^{2}+\right.\right.\nonumber\\
&&\left.\left. \sum\limits_{\circ sites}
\left( \frac{1}{\cosh ^{2} \left( js_{\bullet }^{+} +h\right) }
\right)  \right] \right\}.
\end{eqnarray}
\section{Numerical results}
In all the simulations, square two-dimensional lattices of various sizes 
with periodic boundary 
conditions are used. After the RG blocking, half the spins go, and 
the original boundary conditions are maintained. The heat-bath algorithm 
is used to obtained configurations that are 
distributed with the required weights. One heat bath sweep involves 
visiting every site in the lattice once. 
\subsection{Autocorrelation}
Two additional benefits arise from the improved method.
The first is that the number of sites to be visited is halved. 
While the expressions to be calculated at the remaining sites turn out
to be far more complicated, the use of table look-up means that evaluating
them need not be computationally more expensive.
The second benefit is that correlation between successive 
configurations and hence the number of sweeps required to decorrelate 
data points is reduced. This correlation is quantified in terms of the 
{\it normalised relaxation function} $\phi _{A} 
\left( t\right) $ for some observed quantity $A$,
\begin{eqnarray}
\phi _{A} \left( t\right) &=&\frac{\left\langle\left\langle A\left( 0\right) A\left(
t\right) \right\rangle\right\rangle -
\left\langle\left\langle A\right\rangle\right\rangle ^{2} }
{\left\langle\left\langle A^{2} \right\rangle\right\rangle
-\left\langle\left\langle A\right\rangle\right\rangle ^{2} } 
\end{eqnarray}

The following graph is typical of the behavior near to criticality 
and demonstrates the improvement which is possible. The observable 
used is the real part of the magnetisation versus the number of 
sweeps.
Table 1 shows the data used in generating Fig. 3.

\begin{center}
{{\bf Figure 3:} Autocorrelation of the real
parts of magnetisation.}
\end{center}

\begin{table}
\begin{center}
\caption{\label{table1}Data for Figure 3.}
\begin{tabular}{|cc|}
\hline
Quantity&Value\\
\hline
\hline
Lattice Size		&	32x32\\
Total Sweeps		&	5,000,000\\
Applied Field, $h$	&	0 + 0i\\
Interaction, $j$	&	0.435 + 0.1i\\
Start			&	Cold\\
Walk			& 	Heatbath\\
\hline
\end{tabular}
\end{center}
\end{table}
\subsection{Improved estimate of $\bf <<\rm sign>>$}
As a test of the improved method, it is compared to the crude 
one along the path $ OX$ in Fig. 1.
\begin{center}
{{\bf Figure 4:} $|<<\rm sign>>|$ vs $Re(j)$, $Im(j)=0.1$, 
$h=0$}
\end{center}
Table 2 shows the data used in generating the remainder of the graphs:\\
$\bullet$ {\it Thermalising sweeps} is the number of sweeps performed
before data is collected.\\
$\bullet$ {\it Data points} is the number of configurations used in a 
measurement.\\
$\bullet$ {\it Sweeps between points} is the number of sweeps performed 
between measurements.
\begin{table}
\begin{center}
\caption{\label{table2}Data for Figs. 4-7.}
\begin{tabular}{|cc|}
\hline
Quantity&Value\\
\hline
\hline
Lattice Size		&	20x20\\
Thermalising Sweeps	&	1000\\
Data Points		&	1000\\
Sweeps Between Points	&	100\\
Applied Field, $h$	&	0 + 0i\\
Interaction, $j$	&	Re(j) + 0.1i\\
Start			&	Cold\\
Walk			& 	Heatbath\\
\hline
\end{tabular}
\end{center}
\end{table}

Note the followings from Fig. 4:\\
$\bullet$ Both methods fail close to the phase boundary around 
${\rm Re} $ ( $j$ )= $0.4$ (actually actually no methods can work  
where $Z$ vanishes) and so results for this region are not presented. 
The important thing is that for a given error, 
the improved method is able to get closer to the phase boundary than the crude one.\\
$\bullet$ In agreement with the analytic consideration of section 2, 
the error bars on 
$\left| <<\rm sign>> \right| $
obtained using the improved method are never worse than for the crude one.\\
$\bullet$ Especially for $0.1<j<0.2$, the improved method has lifted
$\left| <<\rm sign>> \right| $ drastically, showing
that $Z\neq 0$ but the crude method cannot.  This results in a big
improvement on the satistical errors of the observables, as we will see
in the following sections. 
The reason that the value of $\left| <<\rm sign>> \right| $ is increased
more at high temperatures is that the summing of opposite spins in
$\cosh \left[ js_{\bullet }^{+} +h\right]$ causes a greater reduction 
in the variance of $e^{-H} /\left| e^{-H} \right|$.

The gains are more striking if we plot the ratio of the proportional errors, 
$R$ where
\begin{eqnarray}
R_{\rm sign}&=&\frac{\left( \frac{S_{\rm sign}}{<<sign>>}\right)_{\rm crude}} 
{\left( \frac{S_{\rm sign}}{<<sign>>}\right)_{\rm improved}}.
\end{eqnarray}
This is an important comparison because
the errors on the physical observables, like magnetisation and susceptibility
$M$ and $\chi$, depend on the proportional error of
$\left| <<\rm sign>> \right|$.

\begin{center}
{{\bf Figure 5:} $R_{\rm sign}$ vs $Re(j)$, 
$Im(j)=0.1$, $h=0$.}
\end{center}
\subsection{Improved estimate of $\bf <M>$}

\noindent
The equivalent graphs for estimates of $\left|<M>\right|$ are presented below.
For clarity, in Fig. 6, only every second data point is shown.
The data agrees with known behaviour of the
magnetisation at high and low temperatures. 
\begin{center}
{{\bf Figure 6:} $|<M>|$ vs $Re(j)$ where 
$Im(j)=0.1$, $h=0$.}
\end{center}
\begin{center}
{{\bf Figure 7:} Ratio of MC error radii of magnetisation.}
\end{center}
\subsection{Improved estimate of susceptibility $\bf \chi$}

\noindent
The magnitudes of susceptibility for crude and improved MC are
shown in Fig. 8.  In the region of OX line in the ferromagnetic
phase, both methods are comparable and consistent with zero.  Fig.
9 depicts the ratio of error radii of the two simulation methods.
\begin{center}
{{\bf Figure 8:} $|<\chi>|$ vs $Re(j)$ where $Im(j)=0.1$, $h=0$.}
\end{center}
\begin{center}
{{\bf Figure 9:} Ratio of MC error radii of susceptibility.}
\end{center}

Comparison with series-expansion data are plotted in Figs. 10 and 11.
\begin{center}
{{\bf Figure 10:} Improved $|<\chi>|$ vs $Re(j)$.  Expansion shown
as line.}
\end{center}

\begin{center}
{{\bf Figure 11:} Improved $<\chi>$ vs $Re(j)$.  Expansion shown as
line.} 
\end{center}
The discs in Fig. 11 are the circles of statistical errors for simulated
results.

\subsection{Dependence on lattice size}
Fig. 5 is actually a slice from Fig 12 and Fig 13. 
These show how the improvement in $R_{sign}$ depends on the linear lattice 
dimensions. Two interesting trends are apparent:\\
$\bullet$ At low temperature the amount of improvement 
increase with lattice size.\\
$\bullet$ At high temperature the amount of improvement 
decrease with lattice size.
We do not attempt to explain this behavior, 
but note that for high temperatures one would expect $R_{sign}$ to approach
some limiting value for large lattice sizes. 
The reason for this is as follows. For both methods, the quantity
$e^{-H} /\left| e^{-H} \right|$ is the sum of $arg$'s over all sites.
Hence, by the central limit theorem for large lattices it is normally distributed. 
Its variance is a function of the spin statistics, which are not size dependent.

{{\bf Figure 12:} $R_{\rm sign}$ vs $Re(j)$ and linear lattice dimension, 
$0.1<Re(j)<0.3$ $Im(j)=0.1$, $h=0$.}

{{\bf Figure 13:} $R_{\rm sign}$ vs $Re(j)$ and linear lattice dimension, 
$0.5<Re(j)<0.8$ $Im(j)=0.1$, $h=0$.}

\subsection{MC renormalisation group}
We explore the MCRG with both the standard and improved methods
at the critical temperature on the positive, real axis.
It is found that the critical exponents of the blocked lattice 
are the same as those on the original.
The values of the critical exponents $\gamma_0$  and $\gamma_1$,
measured using MCRG are displayed in Table 3.
The exact values  of $8/15$ and $1$ are shown at the top of the table.
\begin{table}
\begin{center}
\caption{\label{table3}Critical exponents.}
\begin{tabular}{|ccccc|}
\hline
&$\gamma_0$ (0.533)&&$\gamma_1$ (1.00)&\\
RG iterations&Crude&Improved&Crude&Improved\\
\hline
\hline
1&	0.532(1)&	0.546(1)&	1.13(4)&	1.08(4)\\
2&	0.536(2)&	0.539(3)&	1.12(5)&	1.04(4)\\
3&	0.535(4)&	0.539(4)&	1.20(7)&	1.10(5)\\
4&	0.535(8)&	0.519(5)&	1.05(7)&	1.11(6)\\
\hline
\end{tabular}
\end{center}
\end{table}

The data used in generating Table 3 is shown in Table 4:\\
$\bullet$ The {\it bootstrap method}~\cite{efron} is used to 
calculate errors on the critical exponents. The number of bootstrap 
samples used, $B$, is $500$. In theory, the limit of $B\rightarrow\infty$
should be taken. In practice it is found that the distribution changes 
little for $B>500$.\\
$\bullet$ The results from the crude and improved methods agree 
within error.\\ 
$\bullet$ The consistent deviation from the exact 
value is in agreement with similar simulations~\cite{Swendsen} and 
can be explained by truncation of the hamiltonian during MCRG and 
finite size effects.\\ 
$\bullet$ No improvement should be expected (nor is it observed) 
as there is no sign problem in this case.
The purpose of these figures is only to demonstrate that the improved
method is adaptable for use in MCRG.

\begin{table}
\begin{center}
\caption{\label{table4}Data for Table 3.}
\begin{tabular}{|cc|}
\hline
Quantity&Value\\
\hline
\hline
Lattice Size		&	64x64\\
Data Points		&	1000\\
Sweeps Between Points	&	8000\\
Applied Field, $h$	&	0.001 + 0i\\
Interaction, $j$	&	0.440687 + 0i\\
Start			&	Cold\\
Walk			& 	Heatbath\\
RG Blockings 		&	5\\
Bootstrap Samples	&	500\\
\hline
\end{tabular}
\end{center}
\end{table}

\section*{Concluding remarks}
We have presented a method towards a partial alleviation of the 
sign problem; it is the earlier proposal in~\cite{kieu} generalised
to include exact RG transformations.  The sign problem is lessened
because of some partial phase cancellation among the
original indefinite or complex-valued measure after an exact RG
transformation.  

A particular RG blocking is chosen for our 
illustrative example of the 2D Ising model with complex-valued measure. 
And this summation over a sublattice is the natural choice
which always exists for short-ranged interactions.
But other choices of RG blocking are feasible and how effective they are
depends on the physics of the problems.  

When the quantity to be averaged is not smooth on the length scale of the 
crude weight function, there is an additional source of systematic error 
in the crude, average-sign method.  The cancellation
in the partial sums may reduce this error by reducing the
difference in length scales of the measured quantities and that of the
sampling weights.

\section*{Acknowledgements}
We are indebted to Robert Shrock for discussions and for providing us
the series-expansion data of the susceptibility; to Andy Rawlinson for
the preparation of some graphs.
One of us, TDK, wants to thank Norman Christ and the Theory Group
at Columbia University for their hospitality during his stay.
The authors also wish to thank the Australian Research Council and
Fulbright Program for financial support.  

\newpage
\begin{center}
\large\bf Table Captions
\end{center}

\noindent
{\bf Table 1:} Data for Figure 2.

\noindent
{\bf Table 2:} Data for Figs. 4-7.

\noindent
{\bf Table 3:} Critical exponents.

\noindent
{\bf Table 4:} Data for Table 3.
\newpage
\begin{center}
{\large\bf Figure Captions}
\end{center}

\noindent
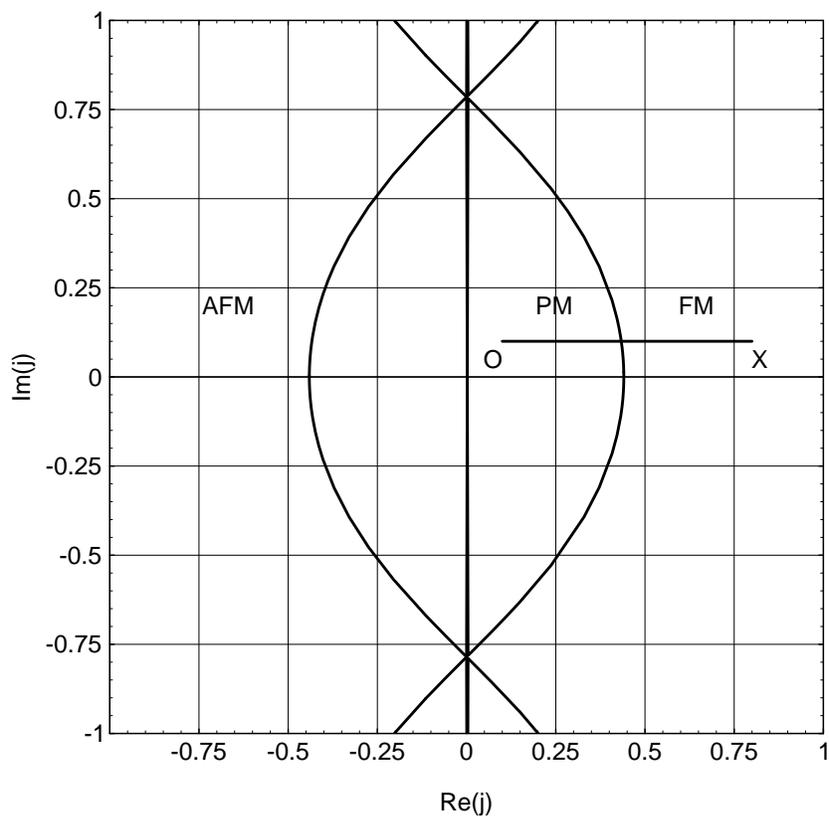
{{\bf Figure 1:} Phase diagram in the complex $j$
plane, $h=0$.  FM=ferromagnetic, PM=paramagnetic, 
AFM=antiferromagnetic.}

\noindent
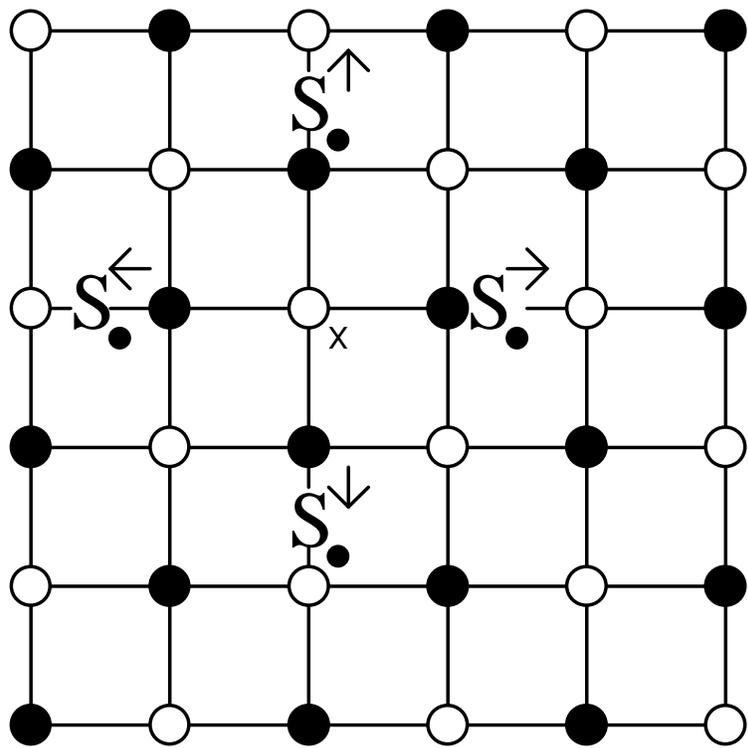
{{\bf Figure 2:} Relative spin positions on a
partitioned lattice.}

\noindent
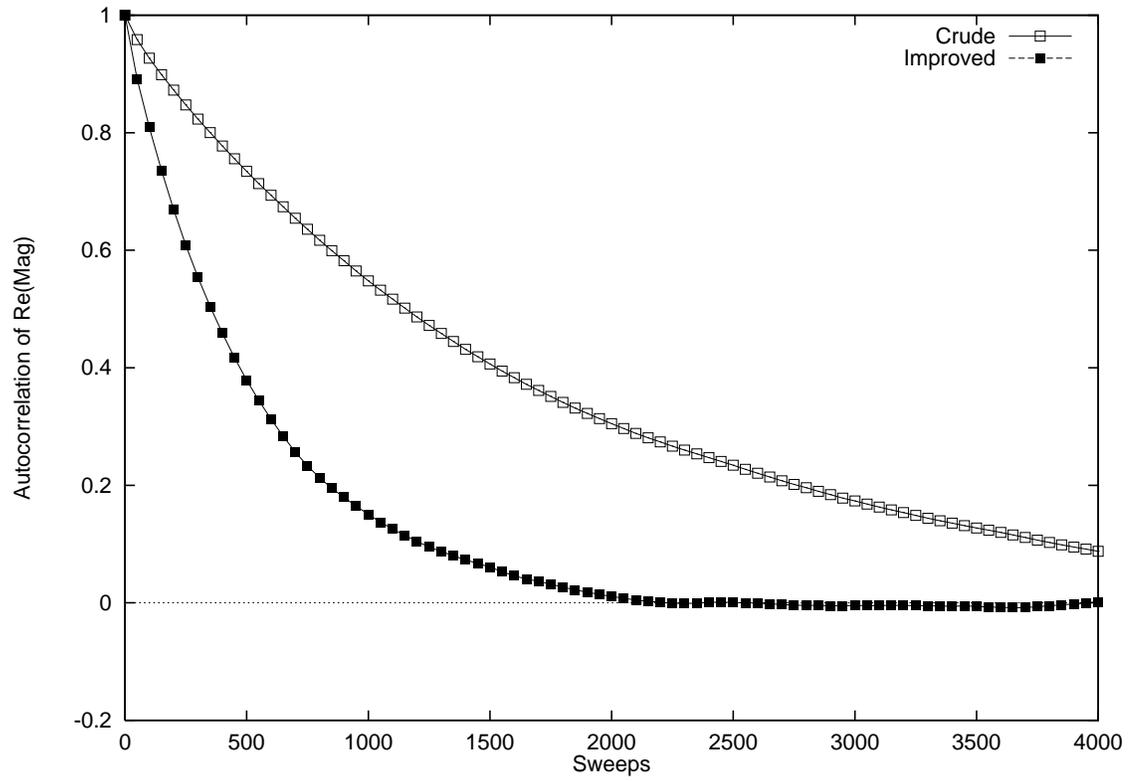
{{\bf Figure 3:} Autocorrelation of the real
parts of magnetisation.}

\noindent
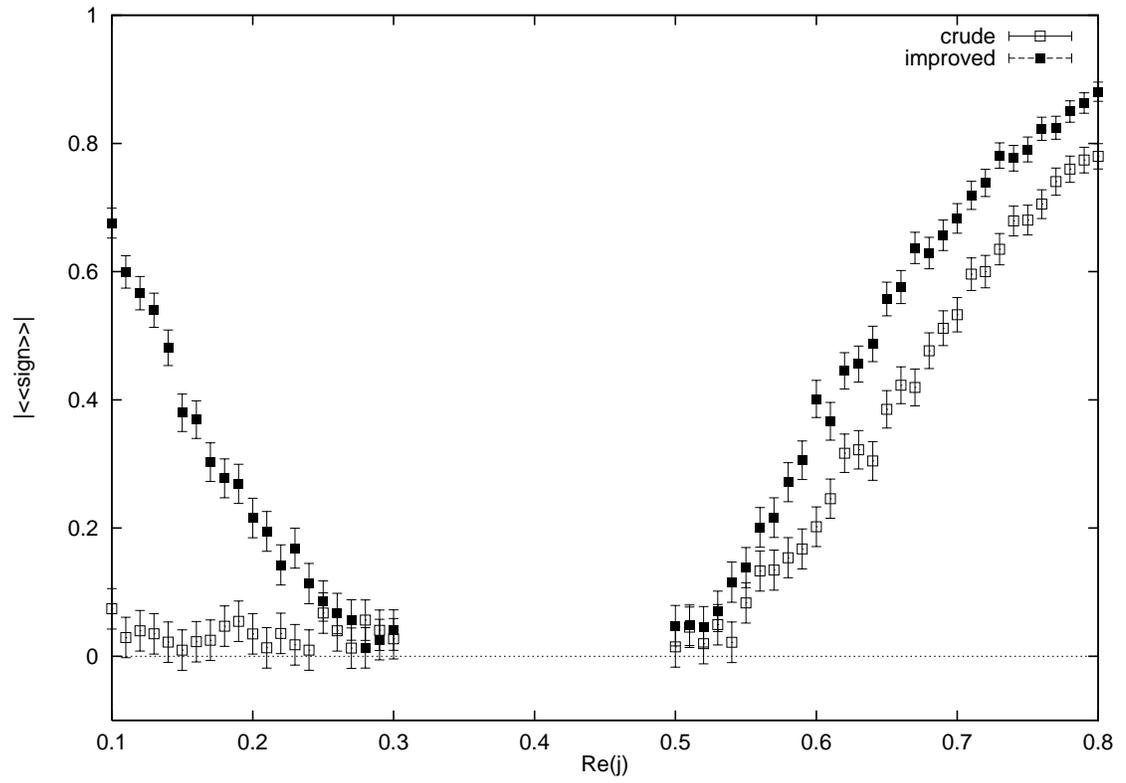
{{\bf Figure 4:} $|<<\rm sign>>|$ vs $Re(j)$, $Im(j)=0.1$,
$h=0$}

\noindent
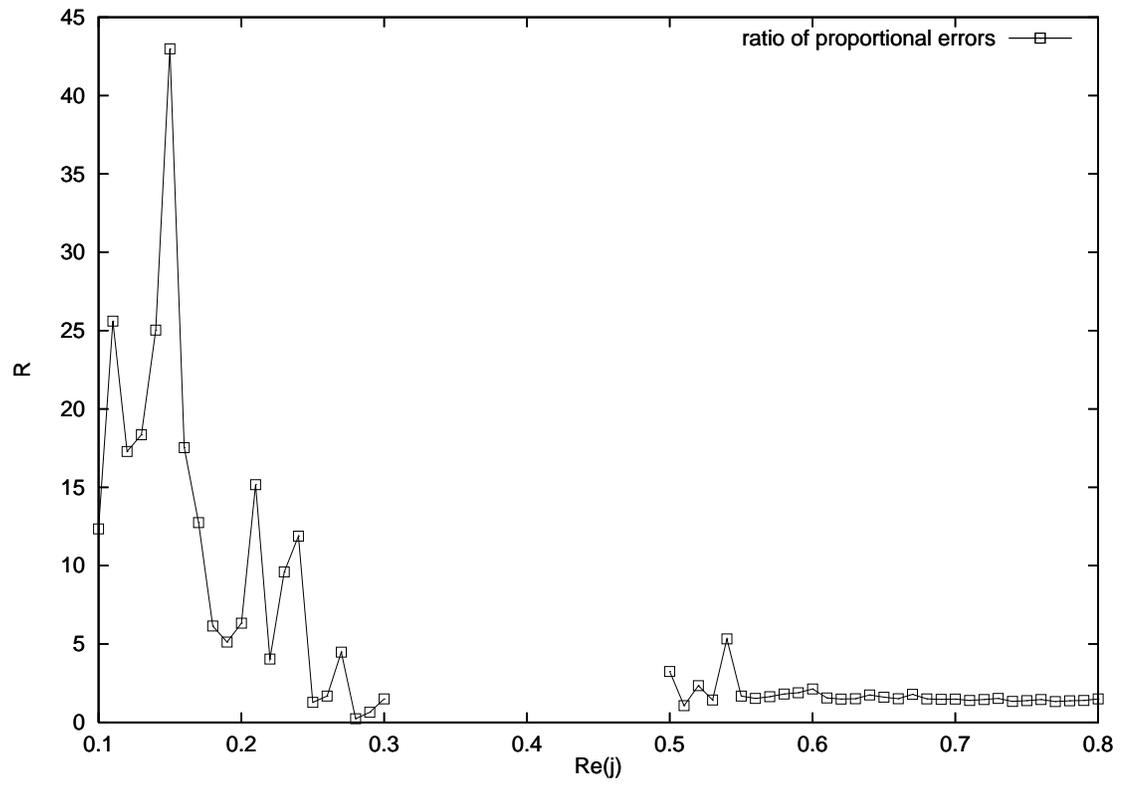
{{\bf Figure 5:} $R_{\rm sign}$ vs $Re(j)$,
$Im(j)=0.1$, $h=0$.}

\noindent
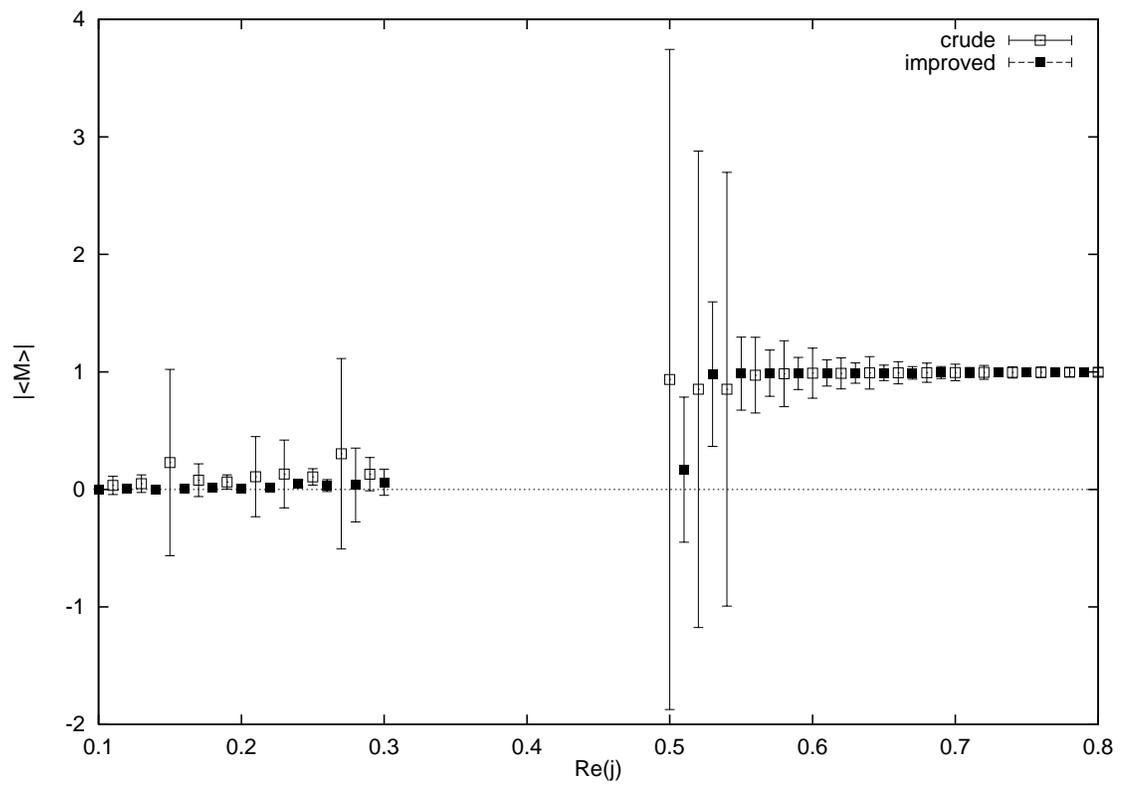
{{\bf Figure 6:} $|<M>|$ vs $Re(j)$ where
$Im(j)=0.1$, $h=0$.}

\noindent
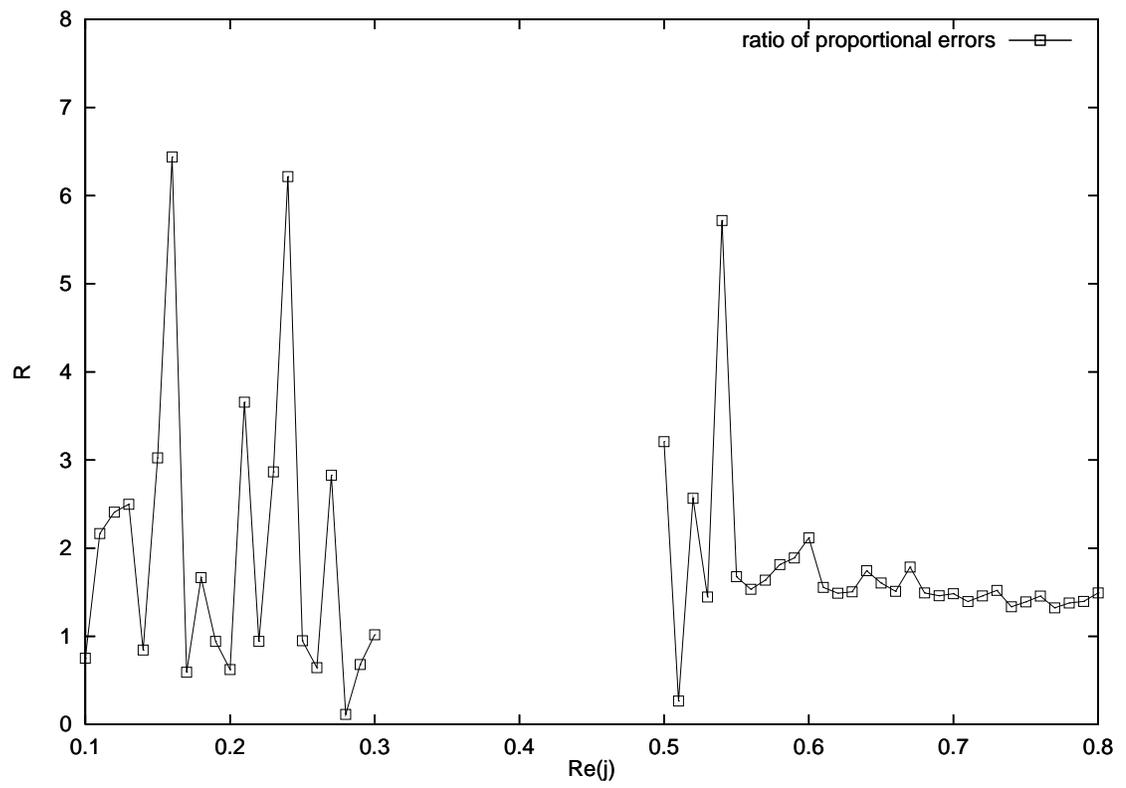
{{\bf Figure 7:} Ratio of MC error radii of magnetisation.}

\noindent
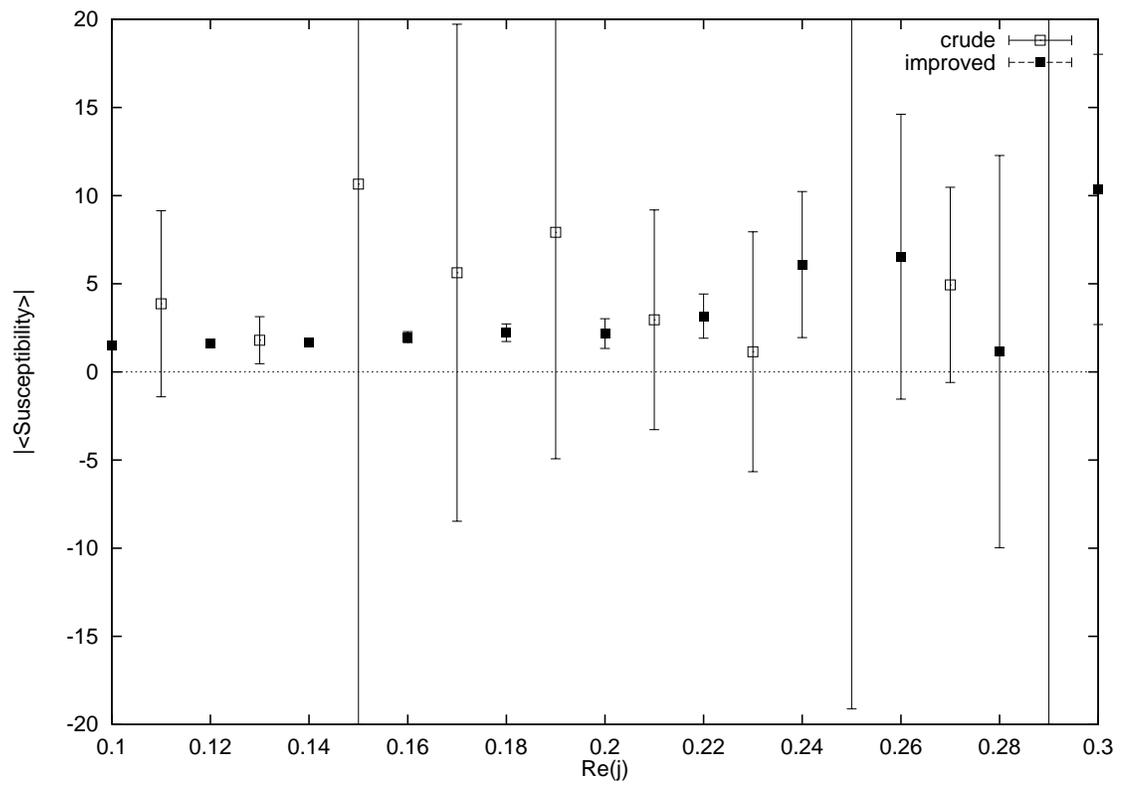
{{\bf Figure 8:} $|<\chi>|$ vs $Re(j)$ where $Im(j)=0.1$, $h=0$.}

\noindent
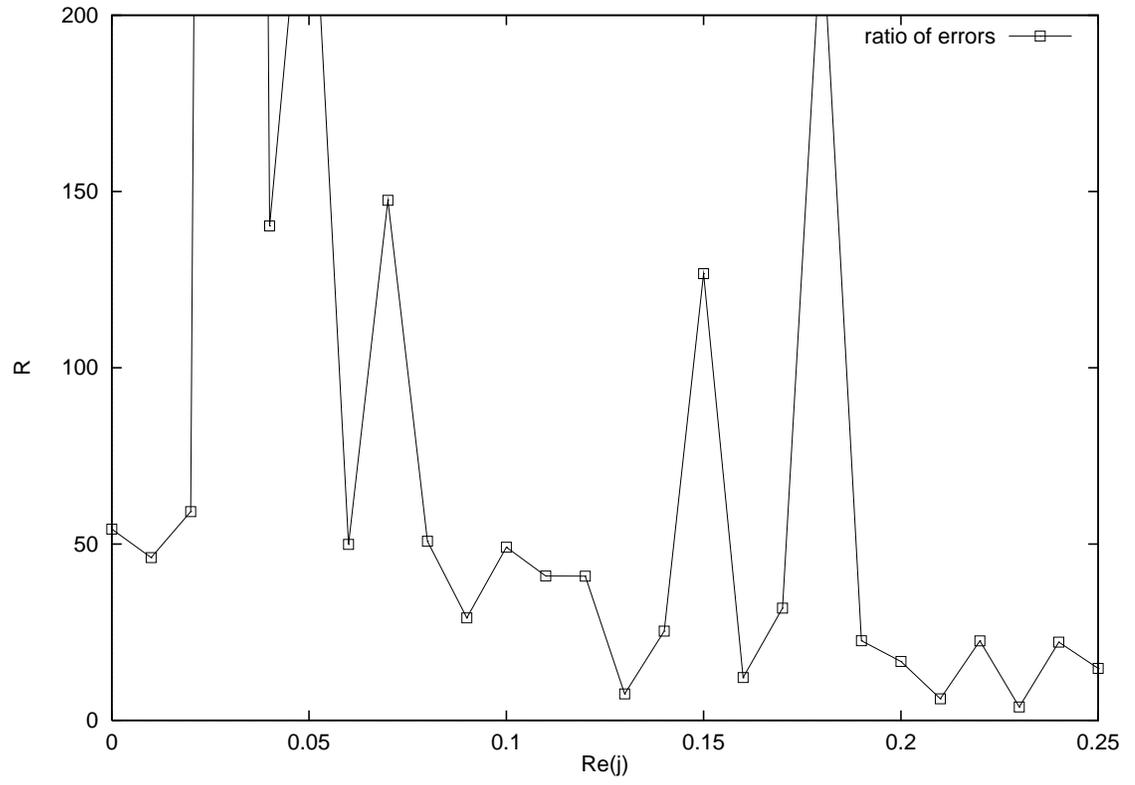
{{\bf Figure 9:} Ratio of MC error radii of susceptibility.}

\noindent
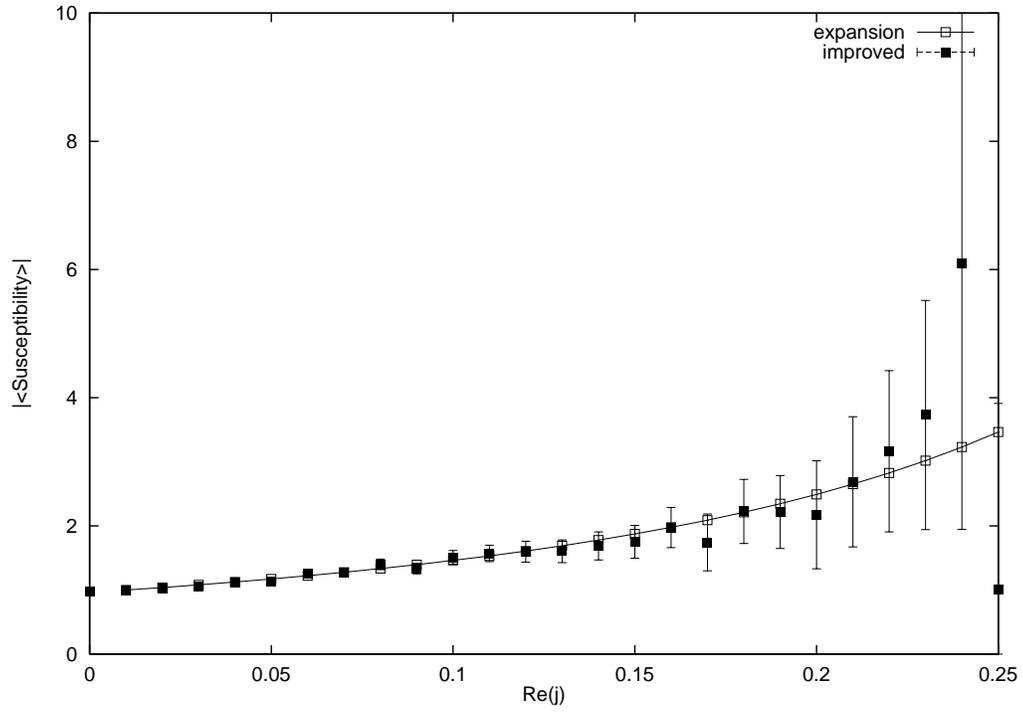
{{\bf Figure 10:} Improved $|<\chi>|$ vs $Re(j)$.  Expansion shown
as line.}

\noindent
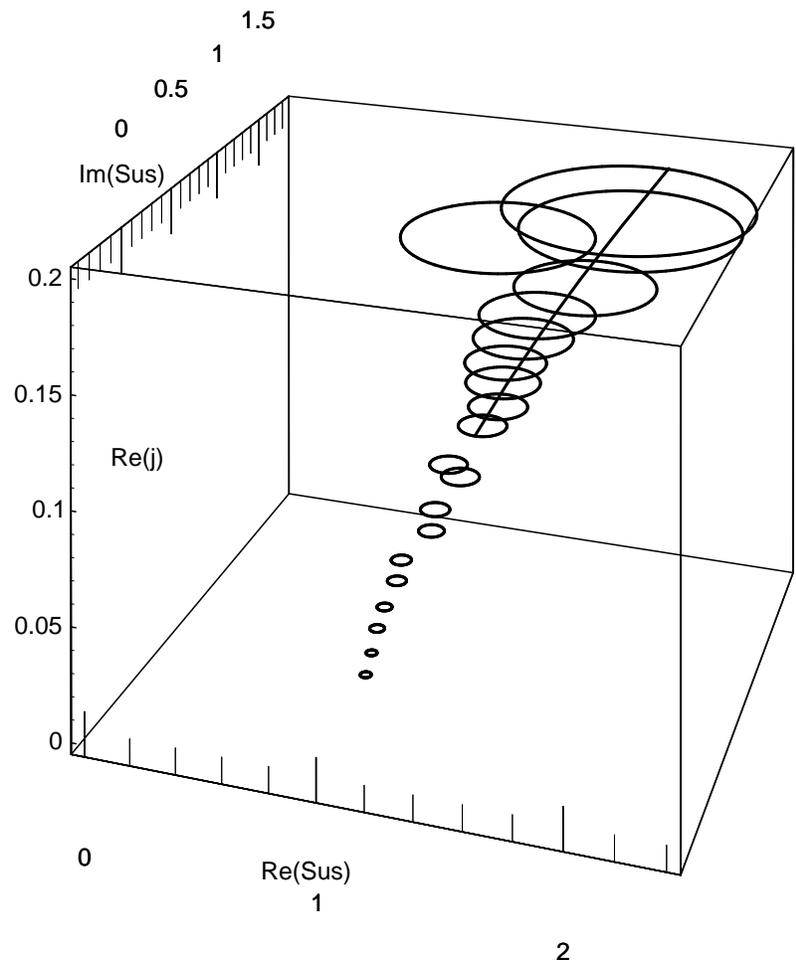
{{\bf Figure 11:} Improved $<\chi>$ vs $Re(j)$.  Expansion shown as
line.}

\noindent
{{\bf Figure 12:} $R_{\rm sign}$ vs $Re(j)$ and linear lattice dimension, 
$0.1<Re(j)<0.3$ $Im(j)=0.1$, $h=0$.}

\noindent
{{\bf Figure 13:} $R_{\rm sign}$ vs $Re(j)$ and linear lattice dimension, 
$0.5<Re(j)<0.8$ $Im(j)=0.1$, $h=0$.}

\end{document}